\documentclass[prl,superscriptaddress,showpacs,amssymb,amsmath,
amsfonts,aps,twocolumn,secnumarabic,floatfix]{revtex4}
\usepackage{graphics}
\usepackage{epsfig}
\begin{document} 
\bibliographystyle{try} 
 
\topmargin 0.1cm 
 
 \title{The Primakoff effect on a proton target}

\newcommand*{\JLAB }{ Thomas Jefferson National Accelerator Facility, Newport News, Virginia 23606} 
\affiliation{\JLAB } 

\author{J.M.~Laget}
     \affiliation{\JLAB}

\date{\today} 
 
\begin{abstract} 

Primakoff effect offers us with a way to determine the radiative decay width of pseudo-scalar mesons when they are photo-produced in the electromagnetic field of hadronic systems. Taking advantage of recent developments in the Regge description of the production of mesons in the strong hadronic field, this paper evaluates the relative importance of the electromagnetic and the strong amplitudes, and  assesses the possibilities which become opened by modern experimental facilities.  
  
\end{abstract} 
 
 
\maketitle

The study of the Primakoff effect~\cite{Pri51} in the coherent photoproduction of pseudo-scalar mesons on nuclear targets has been recently completed at JLab~\cite{ProXY} and will greatly benefit of its energy upgrade to 12 GeV~\cite{pCDR}. The production of a meson in the Coulomb field of the nuclear target gives a direct access to its radiative decay. A precise knowledge of the $\pi^{\circ}$  meson radiative decay provides an accurate test of chiral anomalies and mixing effects due isospin breaking by the difference of the masses of light quarks~\cite{Go02}. The precise measurement of the radiative widths of the $\eta$ and $\eta'$ mesons will lead to an absolute determination of their other decay widths, which are usually measured relative to the radiative decay width. A better understanding  of the  ($\pi^{\circ}$, $\eta$ and $\eta'$) system will result in the determination  of the mixing angles which quantify isospin and SU(3) symmetry breaking~\cite{Go02}. 

However, the mesons can be concurrently photoproduced in the strong nuclear field. This hadronic interaction is mediated by the exchange of vector mesons ($\rho$ and $\omega$) whose singularity is more distant to the physical region than the singularity of the photon which is exchanged in the Primakoff amplitude. Consequently, the strong amplitude contributes to large production angles, but its tail must be subtracted from the Primakoff amplitude which contributes at smaller angles. In previous pioneering experiments~\cite{Bro74a,Bro74b}, this has been achieved by parameterizing the experimental yield at large angles and extrapolating it below the Primakoff peak. While this procedure led to a radiative decay width of the $\pi^{\circ}$ meson in good agreement with the value deduced from collider experiments, it led to a radiative width of the $\eta$ meson about two times smaller~\cite{PDG}.

The quality of the beam of CEBAF at JLab, together with the improvement in the experimental set-up, permit the study of the Primakoff production of pseudo-scalar mesons, and the determination of their radiative decay widths, with an unprecedented statistical, as well as systematic, accuracy. Mastering the hadronic contribution becomes mandatory. This note is an attempt to determine the relative importance and the interplay between the Primakoff and the strong amplitude on the simplest target, the proton. It takes advantage of the latest developments~\cite{Gui97} in the Regge description of the photoproduction of $\pi^{\circ}$ and its extension to the $\eta$ and $\eta'$ sectors.  

\begin{figure}[hbt]
\begin{center}
\epsfig{file=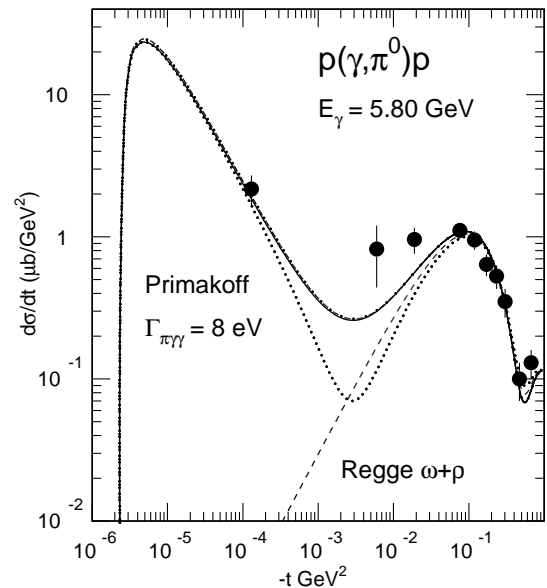,width=3.0in}
\caption[]{The angular distribution of the $\pi^{\circ}$ mesons photoproduced in the reaction $p(\gamma,\pi^{\circ})p$, at $E_{\gamma}=$~5.8 GeV. The dashed curve corresponds to the hadronic cross section, while the full curve corresponds to the coherent sum of the electromagnetic and the hadronic amplitudes. The sign of the electromagnetic amplitude has been changed in the dotted curve.}
\label{pizero}
\end{center}
\end{figure}

Let us start with the $\pi^{\circ}$ production channel. Figure~\ref{pizero} summarizes the results at $E_{\gamma}=$~5.8 GeV, where experimental data~\cite{Bra68,Bra70} exist. The cross sections are plotted against the squared four-momentum transfer $t= (k_{\gamma}-p_{\pi^{\circ}})^2$ between the incident photon ($k_{\gamma}=(E_{\gamma},\vec{k_{\gamma}})$) and the outgoing meson ($p_{\pi^{\circ}}=(E_{\pi^{\circ}},\vec{p_{\pi^{\circ}}})$). Since the energy is large and the mass of the $\pi^{\circ}$ is small, the minimum value, $t_{min}$, of $-t$ is very small and the Primakoff peak is well separated from the hadronic peak. The experimental data seem to prefer a constructive (full line) rather a destructive (dotted) interference between the two contributions, but the tail of the hadronic amplitude does not affect at all the Primakoff peak in the range of small $-t$ where it dominates.

Quantitatively, the hadronic amplitude is the same as in ref.~\cite{Gui97}, where its detailed expression and an extended discussion on the choice of the vertices and coupling constants can be found. Suffice to say that the amplitude is based on the exchange of the Regge trajectories of the $\rho$ and $\omega$ mesons, and takes into account the full spin-isospin structure of the electromagnetic and the strong vertices.  We have chosen a degenerate Regge propagator for the $\omega$, in order to accommodate for the minimum in the experimental angular distribution around $-t= 0.5$~GeV$^2$:
\begin{eqnarray}
\cal{P}_{\omega}&=& (g^{\mu\nu} -\frac{k_{\omega}^{\mu}k_{\omega}^{\nu}}{m^2_{\omega}})
\left( \frac{s}{s_{\circ}}\right)^{\alpha_{\omega}(t)-1}
\frac{\pi \alpha'_{\omega}}{\sin[\pi \alpha_{\omega}(t)]}
\frac{1}{\Gamma(\alpha_{\omega}(t))}
\nonumber \\
& & \times \frac{1 + \exp[-i\pi\alpha_{\omega}(t)]}{2}
 \label{om_prop}
\end{eqnarray}
where the $\omega$ Regge trajectory is:
\begin{eqnarray}
\alpha_{\omega}(t)=0.44+0.9t
 \label{om_traj}
\end{eqnarray}
and $s_{\circ}=1$~GeV is a mass scale. The Gamma function $\Gamma(\alpha(t))$ suppresses the singularities in the physical region ($t<0$). The strong coupling constants are: $g^2_{\omega NN}/4\pi = 17.9$ and $\kappa_{\omega}=0$.

For the $\rho$, we chose a degenerate propagator, with a rotating phase:
\begin{eqnarray}
\cal{P}_{\rho}&=& (g^{\mu\nu} -\frac{k_{\rho}^{\mu}k_{\rho}^{\nu}}{m^2_{\rho}})
\left( \frac{s}{s_{\circ}}\right)^{\alpha_{\omega}(t)-1}
\frac{\pi \alpha'_{\rho}}{\sin[\pi \alpha_{\rho}(t)]}
\frac{1}{\Gamma(\alpha_{\rho}(t))}
\nonumber \\
& & \times \exp[-i\pi\alpha_{\rho}(t)]
 \label{rho_prop}
\end{eqnarray}
where the $\rho$ Regge trajectory is:
\begin{eqnarray}
\alpha_{\rho}(t)=0.55+0.8t
 \label{fho_traj}
\end{eqnarray}
The strong coupling constants are: $g^2_{\rho NN}/4\pi = 0.92$ and $\kappa_{\rho}=6.1$.

The coupling constant of the electromagnetic vertex (as defined in~\cite{Gui97}) is related to the corresponding decay width of the vector meson through the expression:
\begin{eqnarray}
\Gamma_{V\rightarrow \pi\gamma}= 
\frac{\alpha_{em}(m^2_V-m^2_{\pi})^3}{24m^3_Vm^2_{\pi}}\; g^2_{V\pi\gamma}
 \label{V_decay}
\end{eqnarray}
This gives $g_{\omega\pi\gamma}= 0.314$, for $\Gamma_{\omega\rightarrow \pi\gamma}=720$~keV, and $g_{\rho\pi\gamma}= 0.103$, for $\Gamma_{\rho\rightarrow \pi\gamma}=68$~keV~\cite{Gui97}.

It turns out that the contraction between the mass dependent term $k_{\rho}^{\mu}k_{\rho}^{\nu}/m^2_{\rho}$ of the vector meson propagator and the electromagnetic vertex vanishes, and only the contribution of the $g^{\mu\nu}$ part survives. Therefore, the Primakoff amplitude takes the same form as the vector meson exchange amplitudes, provided that the Regge propagator is replaced by the Feynman propagator of the photon
\begin{eqnarray}
{\cal P}_{\gamma}= \frac{g^{\mu\nu}}{t} 
 \label{gam_prop}
\end{eqnarray}
and the strong coupling constants by $g^2_{\gamma NN}/4\pi = \alpha_{em}$ and $\kappa_{\gamma}=1.79$. The Primakoff coupling constant is related to the two photon decay width as follows
\begin{eqnarray}
\Gamma_{\pi\rightarrow \gamma\gamma}= 
\frac{\alpha_{em} m_{\pi}}{16}\; g^2_{\pi\gamma\gamma}
 \label{Prim_decay}
\end{eqnarray}

The full curve in Figure~\ref{pizero} corresponds to the central value $\Gamma_{\pi\rightarrow \gamma\gamma}=8$~eV of ref.~\cite{PDG} ($g_{\pi\gamma\gamma}=0.0114$), while the dot-dashed curve uses 8.4~eV at the edge of the experimental values.

Due to the structure of the electromagnetic vertices, both the hadronic amplitude and the Primakoff amplitudes strickly vanish at $\theta_{\pi}=0$, and therefore $t_{min}$. However, the photon pole is so close to the physical region that it boosts the Primakoff amplitude orders of magnitude above the strong amplitude.

$b_1$ meson exchange may also contribute to the strong amplitude. In ref.~\cite{Gui97} we found that, while it is necessary to reproduce the beam asymmetry, it only modifies by about 10\% the unpolarized cross section in the region of the minimum ($-t\sim 0.5$~ GeV$^2$) and above. So it may modify by the same amount the strong amplitude below $-t\sim 10^{-3}$~GeV$^2$. Since our knowledge of its coupling constants is on less solid ground than those of $\rho$ and $\omega$ mesons, I do not retain its contribution.

The model confirms the earlier estimate of the cross section which was included in the experimental papers~\cite{Bra68,Bra70}. While there is no reason why the Primakoff amplitudes be different, the strong amplitude included exchanges of the $\omega$ and $b_1$ meson Regge trajectories and the $\omega$~Pomeron cut, following refs.~\cite{Ad67,Ca69}. The residues of the poles were fitted to the data, contrary to the model which I use: in ref.~\cite{Gui97} we chose the values of strong and the electromagnetic coupling constants  in the range of values determined in independent channels, and we implemented the full spin-isospin structure of the lower mass realization of each Regge trajectory. 

In Figure~\ref{pizero} the experimental points have been plotted at the mean value of $-t$ in each experimental bin in $\theta_{\pi^{\circ}}$. A new experimental determination of the $\pi^{\circ}$ Primakoff cross section, with a better angular resolution, is highly desirable, especially in the domain where the Primakoff and the strong amplitudes interfere.  

\begin{figure}[hbt]
\begin{center}
\epsfig{file=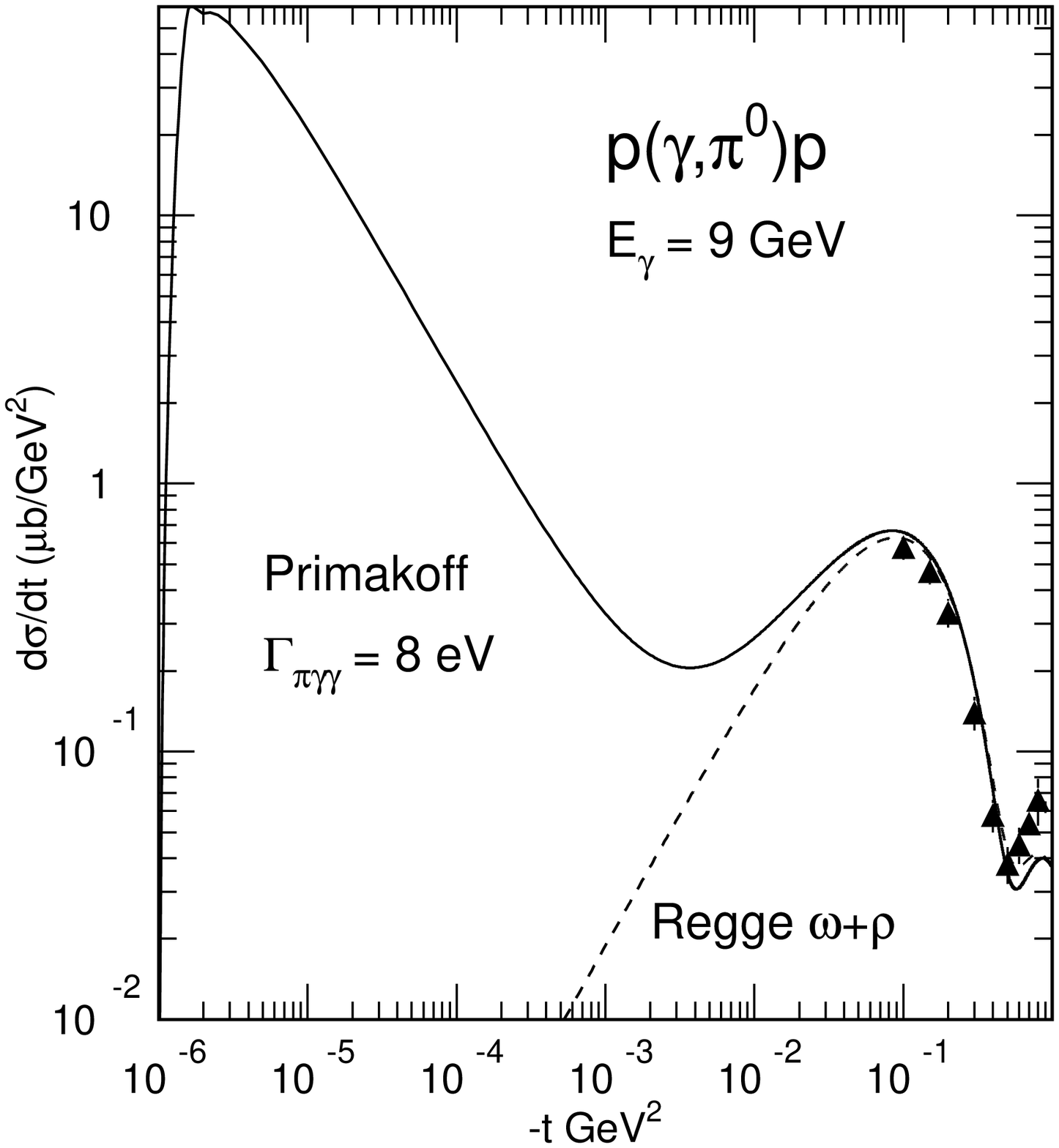,width=3.0in}
\caption[]{The angular distribution of the $\pi^{\circ}$ mesons photoproduced in the reaction $p(\gamma,\pi^{\circ})p$, at $E_{\gamma}=$~9 GeV. The meaning of the curves is the same as in Fig.~\ref{pizero}.}
\label{pizero_9}
\end{center}
\end{figure}

Increasing the energy up to $E_{\gamma}=9$~GeV does not change dramatically the picture. As it can be seen in Figure~\ref{pizero_9}, $t_{min}$ decreases by a factor two, but this does  not improve the separation between the Primakoff and the hadronic peak, which was already very good at 5.8~GeV. Note that the Regge model reproduces quite well the experimental data~\cite{An70} at this energy also.

\begin{figure}[hbt]
\begin{center}
\epsfig{file=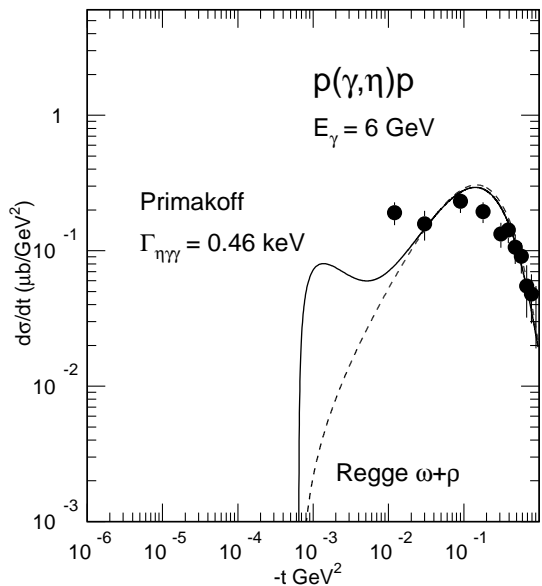,width=3.0in}
\caption[]{The angular distribution of the $\eta$ mesons photoproduced in the reaction $p(\gamma,\eta)p$, at $E_{\gamma}=$~6 GeV. The meaning of the curves is the same as in Fig.~\ref{pizero}.}
\label{eta_6}
\end{center}
\end{figure}

On the contrary, increasing the energy helps in the $\eta$ and $\eta'$ channels. Figure~\ref{eta_6} shows the angular distribution of the $\eta$ emitted in the $p(\gamma,\eta)p$ reaction at $E_{\gamma}= 6$~GeV. The value of $t_{min}$
is more than two orders of magnitude bigger than in the $\pi^{\circ}$ channel. Consequently, one has to rely on the tail of the Primakoff amplitude, which shows up above the tail of the strong amplitude.

The model uses the same $\rho$ exchange amplitude (degenerate Regge propagator, eqs.~\ref{rho_prop}, with the same strong coupling constants) as in the $\pi^{\circ}$ channel. The radiative coupling constant  $g_{\rho\eta\gamma}= 0.81$ is deduced from the experimental decay width $\Gamma_{\rho\rightarrow \eta\gamma}=39$~keV~\cite{PDG} with eq.~\ref{V_decay} (where the $\pi$ mass is replaced by the $\eta$ mass, $m_{\eta}$).

In the $\omega$ exchange amplitude, the radiative decay constant  $g_{\omega\eta\gamma}= 0.29$ is also fixed by the experimental decay width $\Gamma_{\omega\rightarrow \eta\gamma}=5.4$~keV~\cite{PDG}. But, following ref.~\cite{Wen03}, I use the degenerate form, eq.~\ref{rho_prop}, instead of the non degenerate form which I used in the $\pi^{\circ}$ channel. The reason is that available experimental data~\cite{Bra70b} do not exhibit a minimum in the vicinity of the first node of the $\omega$ non-degenerate Regge trajectory. In nature, Regge trajectories (in the form of eq.~\ref{om_prop}) go by pair, each having the same slope but a different signature, ${S}=\pm 1$, when it connects members with either odd or even spins. When it happens that each trajectory has the same, or comparable, coupling constants with the probe, they combine into a degenerate trajectory with or without a rotating phase (eq.~\ref{rho_prop}). Only experiment tells us what is the best choice. The conjecture is that the photon couples to a degenerate trajectory of the $\omega$ in the $\eta$ photoproduction channel, while it couples to a non-degenerate $\omega$ trajectory in the $\pi^{\circ}$ sector. Consequently the strong coupling constants of the omega are not necessarily the same in both channels. A good agreement with the data is achieved when I use $g^2_{\omega NN}/4\pi = 6.44$ and I keep $\kappa_{\omega}=0$.

This set of coupling constants is different from the set of ref.~\cite{Wen03} which has been obtained from a global fit in the $\eta$ and $\eta'$ sector. I prefer to use a set which differs in a minimal way from the $\pi^{\circ}$ sector set. It is worth to note that both sets lead to a similar accounting of the strong part of the cross section.

The Primakoff coupling constant of the $\eta$ meson, $g_{\eta\gamma\gamma} = 0.0429$, is deduced from
the average (including Primakoff measurement) value $\Gamma_{\eta\rightarrow \gamma\gamma}=0.46$~keV of ref.~\cite{PDG} with the help of eq.~\ref{Prim_decay} (where the $\pi$ meson mass is replaced by the $\eta$ meson mass).

\begin{figure}[hbt]
\begin{center}
\epsfig{file=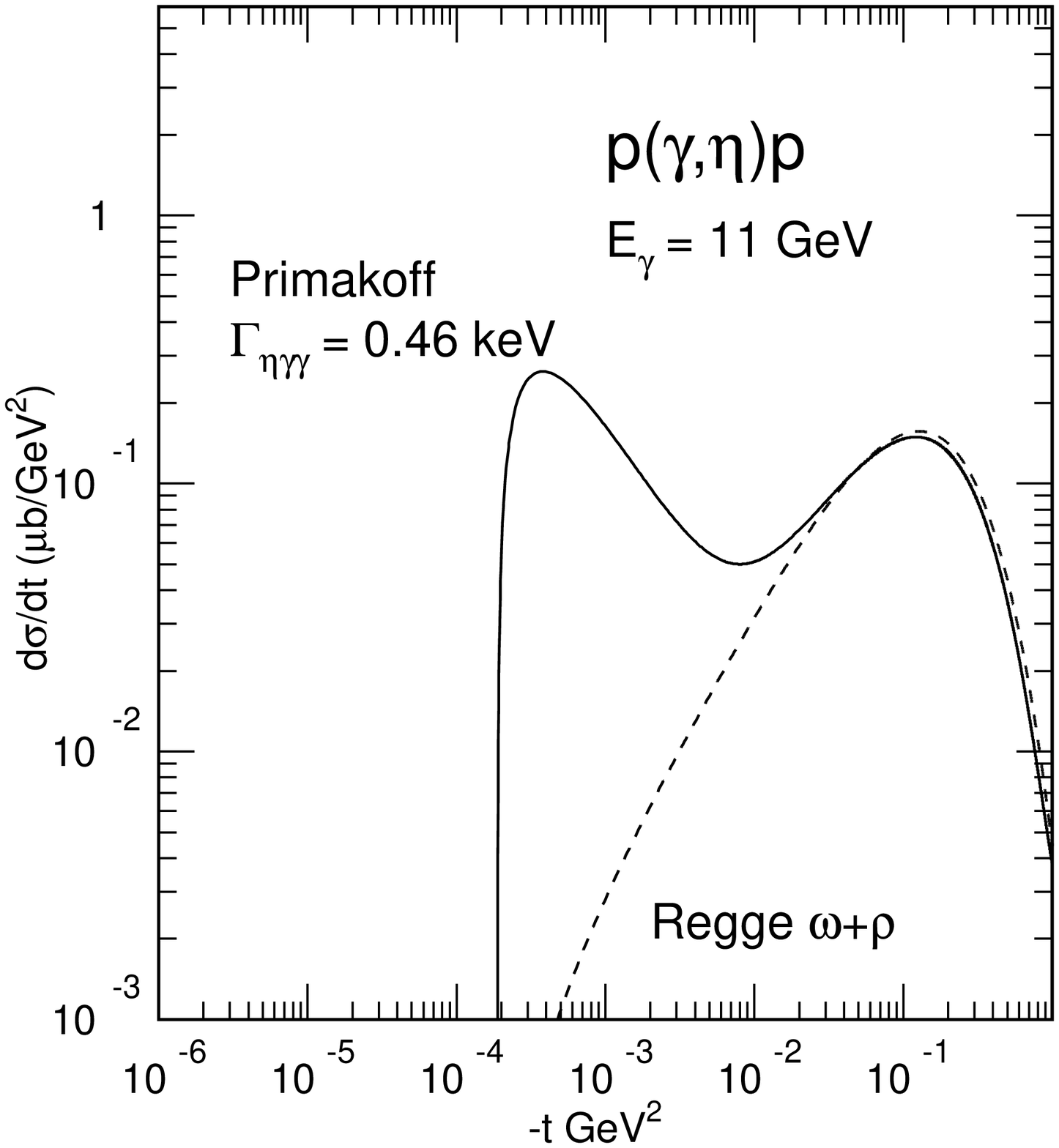,width=3.0in}
\caption[]{The angular distribution of the $\eta$ mesons photoproduced in the reaction $p(\gamma,\eta)p$, at $E_{\gamma}=$~11 GeV. The meaning of the curves is the same as in Fig.~\ref{pizero}.}
\label{eta_11}
\end{center}
\end{figure}

The experimental study of the Primakoff effect will greatly benefit of the increase of the incoming photon energy. Figure~\ref{eta_11} clearly demonstrates that, at $E_{\gamma}= 11$~GeV, $t_{min}$ is lowered by about a factor three and the Primakoff peak is clearly separated from the strong hadronic peak.

Let us turn now to the $\eta'$ channel, where I keep the same amplitudes as in the $\eta$ channel and only change the radiative coupling constants. The $\eta'$ radiative decay to the $\rho$ or the $\omega$ mesons is related to the radiative coupling constant in the following way:
\begin{eqnarray}
\Gamma_{\eta'\rightarrow V\gamma}= 
\frac{\alpha_{em}(m^2_{\eta'}-m^2_V)^3}{8m^5_{\eta'}}\; g^2_{V\eta'\gamma}
 \label{etaprim_decay}
\end{eqnarray} 
This corresponds to $g_{\omega\eta'\gamma}= 0.43$, for $\Gamma_{\eta'\rightarrow \omega\gamma}=6$~keV, and $g_{\rho\eta'\gamma}= 1.24$, for $\Gamma_{\eta'\rightarrow \rho\gamma}=55$~keV.

Again the Primakoff coupling constant of the $\eta'$ meson, $g_{\eta'\gamma\gamma}= 0.0989$, is deduced from
the central value $\Gamma_{\eta'\rightarrow \gamma\gamma}=4.27$~keV of ref.~\cite{PDG} with the help of eq.~\ref{Prim_decay} (where the $\pi$ meson mass is replaced by the $\eta'$ meson mass).

\begin{figure}[hbt]
\begin{center}
\epsfig{file=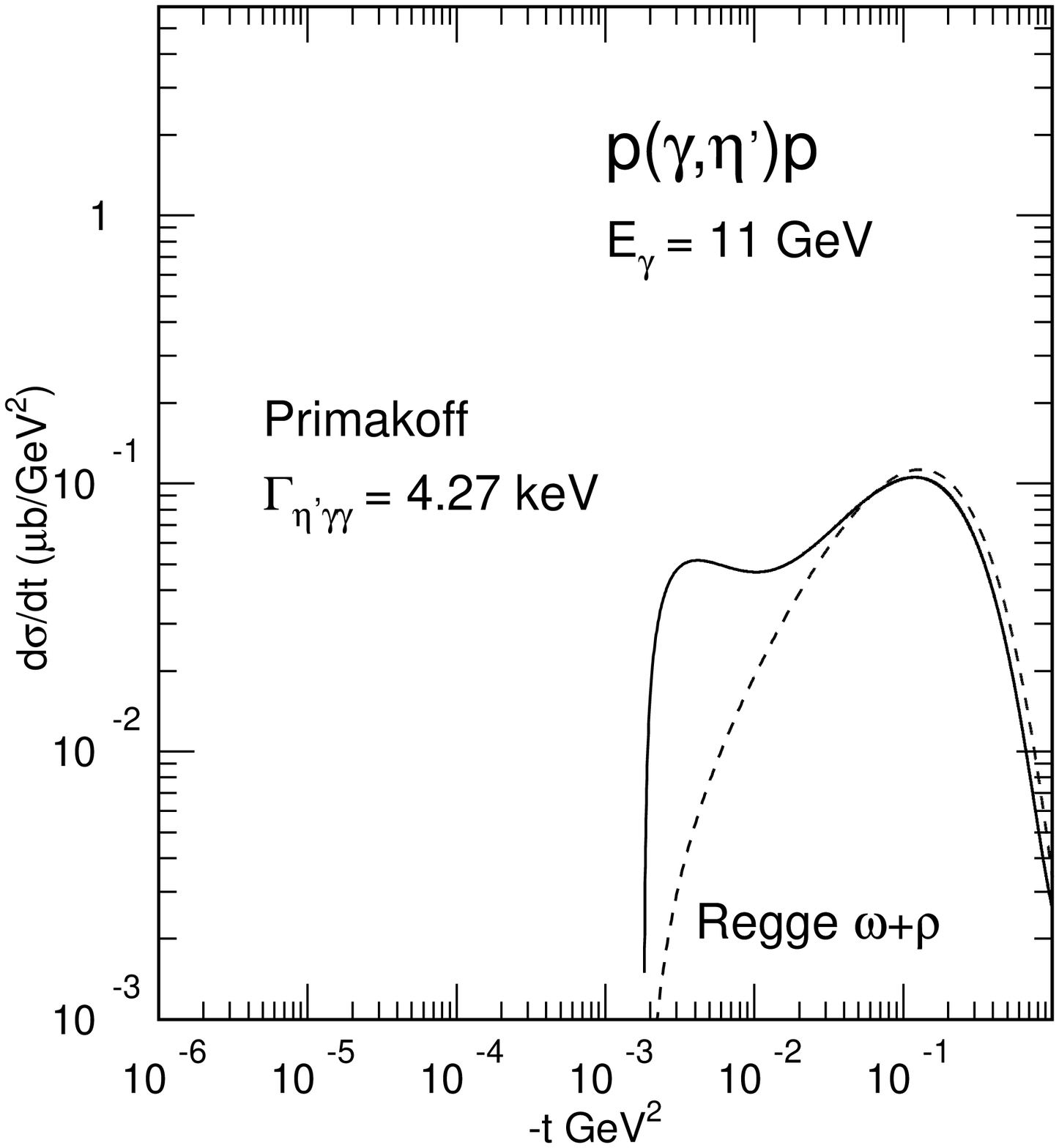,width=3.0in}
\caption[]{The angular distribution of the $\eta'$ mesons photoproduced in the reaction $p(\gamma,\eta')p$, at $E_{\gamma}=$~11 GeV. The meaning of the curves is the same as in Fig.~\ref{pizero}.}
\label{etaprim}
\end{center}
\end{figure}

The results are shown in Figure~\ref{etaprim} at $E_{\gamma}=11$~GeV. Even at such a high energy, $t_{min}$ is high and the Primakoff contribution appears as  a shoulder on the tail of the strong amplitude. The situation is similar as in the $\eta$ sector at $E_{\gamma}=6$~GeV. Certainly an experiment with an excellent energy resolution will disentangle the Primakoff amplitude at the lowest angles, where it overwhelms by more that an order of magnitude the strong amplitude, which will be calibrated at higher angles. A measurement of $\eta$ meson production at $E_{\gamma}=6$~GeV would be very welcome in this respect: it is already possible.

Contrary to the $\pi^{\circ}$ and the $\eta$ channels, the experimental data set is extremely scarce. The model can only be compared to integrated experimental cross sections~\cite{ABBHHM,Str76}. At $E_{\gamma}=5$~GeV, it predicts $\sigma=0.1$~$\mu$b in the range of the experimental cross section $\sigma_{exp}= 0.17\pm 0.12$~$\mu$b. This gives confidence in its extrapolation from the $\eta$ to the $\eta'$ channel. But a more accurate determination of the angular distribution is definitely needed.

In conclusion, the Regge description of the strong hadronic amplitude has been extended from $\pi^{\circ}$ to $\eta$ and $\eta'$ photo-production. It reproduces all the available experimental data and provides us with a solid starting point to evaluate the hadronic contribution below the Primakoff peak. The Primakoff amplitude has been incorporated in the model. While in the $\pi^{\circ}$ sector it is prominent already at $E_{\gamma}=6$~GeV, its determination in the $\eta$ and $\eta'$ sector requires accurate experiments at higher energies.

I would like to thank A. Bernstein and R. Miskimen who triggered my interest in the Primakoff effect, J. Goity who taught me its relevance to QCD and M. Vanderhaeghen for discussions on Regge models.

\end{document}